\documentstyle{l-aa}
\begin{document}
\title{
The Optical Gravitational Lensing Experiment.
Variable stars in globular clusters -IV. Fields 104A-E in 
47~Tuc
\thanks{
Based on observations collected at the Las Campanas Observatory of the
Carnegie Institution of Washington.}
}
\author{J. \, Kaluzny\inst{1} \and M. \, Kubiak\inst{1} \and
M. \, Szyma{\'n}ski\inst{1} \and A. \, Udalski\inst{1} \and
W. \, Krzemi{\'n}ski\inst{2} \and Mario Mateo\inst{3} \and
K.Z. \, Stanek\inst{4}}

\offprints{J. Kaluzny}

\institute{Warsaw University Observatory, Al. Ujazdowskie ~4, 
00-478 Warsaw, Poland,
e-mail: (jka,mk,msz,udalski)@sirius.astrouw.edu.pl
\and
Carnegie Observatories, Las Campanas Observatory, Casilla 601,
La~Serena, Chile,
e-mail: wojtek@roses.ctio.noao.edu
\and
Department of Astronomy, University of Michigan, 821 Dennison
Bldg., Ann Arbor, MI 48109-1090, USA,\\
e-mail: mateo@astro.lsa.umich.edu
\and 
Harvard-Smithsonian Center for Astrophysics, 60 Garden St., Cambridge,
MA 02138, USA,\\ 
e-mail: kstanek@cfa.harvard.edu
}
\date{Received\dots, accepted\dots}
\maketitle
\markboth{Kaluzny et al.: Variable stars in 47~Tuc}{}
\begin{abstract}
Five fields located close to the center of the globular cluster 
NGC~104=47~Tuc were surveyed in a search for variable stars.
We present $V$-band light curves for 42 variables. This sample
includes 13 RR~Lyr stars -- 12 of them belong to the Small Magellanic 
Cloud (SMC) and 1
is a background object from the galactic halo. 
Twelve eclipsing binaries were identified -- 9 contact systems and 3
detached/semi-detached systems. Seven eclipsing binaries are located
in the blue straggler region on the cluster color-magnitude diagram (CMD)  
and four binaries
can be considered main-sequence systems.  One binary is probably a
member of the SMC. 
Eight contact binaries are likely members of the cluster
and one is most probably a foreground star. We show that for the
surveyed region of 47~Tuc, the relative frequency of contact binaries
is very low as compared with other recently surveyed globular clusters.
The sample of identified variables also includes 15 red variables 
with periods ranging from about 2 days to several weeks. 
A large fraction of these 15 variables probably belong to the SMC
but a few stars are likely to be red giants in 47~Tuc. 
$VI$ photometry for about 50 000 stars from the cluster fields
was obtained as a by product of our survey.
\footnote{
The photometric data presented in this  paper
are available in electronic form  at the CDS, via ftp 130.79.128.5}
\end{abstract}
\section{Introduction}
The Optical Gravitational Lensing Experiment (OGLE) is a long term
project with the main goal of searching for dark matter in our Galaxy by
identifying microlensing events toward the galactic bulge (Udalski et
al. 1992, 1994). At times  the bulge is unobservable we conduct other
long-term photometric programs.
A complete list of side-projects attempted by the OGLE team
can be found in Paczy{\'n}ski et al. (1995).
In particular, during the observing seasons 1993, 1994 and 1995 we 
monitored globular clusters NGC~104=47~Tuc and  NGC~5139=$\omega$~Cen
in a search for
variable stars of various types. Of primary interest was the detection of 
detached eclipsing binaries. In Papers I, II \& III  (Kaluzny et al.
1996, 1997a, 1997b) we
presented results for $\omega$~Cen. Here we report on
variables discovered in the field of 47~Tuc.

\section{Observations and data reduction}

The OGLE\footnote {The OGLE project is currently conducted,
under the name OGLE-2, using a dedicated 1.3-m telescope 
located at Las Campanas Observatory}
project was conducted using the 1-m Swope telescope at Las
Campanas Observatory.
A single $2048\times 2048$ pixels Loral CCD chip, giving a
scale of 0.435 arcsec/pixel was used as the detector. The initial
processing of the raw frames was done automatically in near-real time.
Details of the standard OGLE processing techniques were described by
Udalski et al. (1992).

In 1993 we monitored fields 104A and 104B located west 
and east of the cluster center, respectively. In 1994 we monitored field
104C located north of the cluster center. In 1995 we
monitored fields 104D and 104E covering southern part of 
the cluster.  
A condensed summary of the data used in this paper is given in
Table 1.
Detailed logs of the observations can be found in Udalski
et al. (1993, 1995, 1997). 
The equatorial coordinates of centers of fields 104A-E are given in
Table 2. 
A schematic chart with marked locations of all of the  
monitored fields is shown in Fig. 1. 
Most of the monitoring was performed through the Johnson $V$ filter. 
Some exposures in the Kron-Cousins $I$ band were also obtained. 
Most of observations in the $V$-band were collected with an exposure time
ranging from 300 to 600  seconds (420 seconds was the most common value).
The $I$-band exposures lasted 300 seconds.  
For the majority of the analyzed frames the seeing was better than 1.6 arcsec. 
The reduction techniques as well as the algorithms used for selecting
potential variables are described in Paper I. Profile photometry 
was extracted with the help of DoPHOT (Schechter et al. 1993). 
The total number of stars contained in data bases with $V$ band photometry
ranged from 18397 to 33014. 
Table 3 gives condensed information about the numbers of stars analyzed for
variability and about the quality of the derived photometry.
The useful data were obtained for stars with $14.0<V<20.25$.
\section{Variable stars}
In this paper we present results for 42 variables identified in
the five observed fields. All except two are new discoveries and 
were assigned names OGLEGC212-255. Names OGLEGC217 and OGLEGC224
were given to previously known variables V9 and V3 (eg. Hogg 1973).
Photometry obtained for these two stars was
poor because their images were badly overexposed on most of
analyzed frames. Therefore, we decided to drop OGLEGC217=V9 and
OGLEGC224=V3 from our list of variables.
 
The rectangular and equatorial coordinates of the 42 newly identified
variables are listed in Table 4\footnote
{In fact variability of 
OGLEGC212, 213, 214, 216, 243, 245 and  246 was reported recently by
Kaluzny et al. (1997c). These authors surveyed a western part of
the cluster covering a region overlapping with fields 104A and 104C.
It is encouraging that all but one variable from Kaluzny et al. (1997c)
were recovered in the current study. We missed variable V8 which
is very faint and was not included on the list of template stars for
the field 104A.}.
The rectangular coordinates correspond to
positions of variables on the $V$-band "template" images. 
These images 
allow easy identification of all objects listed in Table 4. 
The name of the field in which a given variable can be 
identified is given in the 6th column. 
All frames collected by the OGLE team were deposited at the NASA NSS
Data Center \footnote{The OGLE data (FITS images) are accessible to
astronomical community
from the NASA NSS Data Center. Send e-mail to:
archives@nssdc.gfc.nasa.gov with the subject line: REQUEST OGLE ALL and 
put requested  frame numbers (in the form MR00NNNN where NNNN stands for
frame number according to OGLE notation), one per line, in the body
of the message. Requested frames will be available using 
an "anonymous ftp" service from nssdc.gfc.nasa.gov host in 
location shown in the return e-mail message from 
archives@nssdc.gsfc.nasa.gov
}. 
Frames mr5228, mr5227, mr7890, mr14597 and mr14595 were used as templates for 
fields 104A, 104B, 104C , 104D and 104E, respectively.   
The transformation from rectangular to equatorial 
coordinates was derived from positions of stars which could be 
matched with objects from the astrometric list kindly provided by 
Kyle Cudworth. The number of "transformation stars" identified in a given
field ranged from 55 to 100.
The adopted frame
solutions reproduce equatorial coordinates of these stars with residuals 
rarely exceeding 0.5 arcsec. According to Cudworth the absolute accuracy
of equatorial coordinates for stars from his table is not worse than
$2\arcsec$. 

Our sample of variables includes 13 RR~Lyr stars. 
Table 5 lists basic characteristics of the light curves of these stars. 
The mean $V$ magnitudes were calculated
by numerically integrating the phased light curves after converting them
into an intensity scale. 
Photometric data for the remaining variables 
are given in Table 6.
The $V-I$ colors listed in Tables 5 and 6 were measured at random 
phases. For each of fields we used a single exposure in the $I$ band
bracketed by two exposures in the $V$ band. 
To determine the periods of identified variables we used an {\it aov} 
statistic
(Schwarzenberg-Czerny 1989, 1991). This statistic allows -- in 
particular -- reliable determination of periods for variables with 
non-sinusoidal light curves (eg. eclipsing binaries).
Phased light curves of RR~Lyr  stars are shown in Figs. 2 \& 3 while 
Fig. 4 presents phased light curves for the remaining variables 
with determined periods. Time domain light curves for these 
variables for which we were unable to determine periods are 
shown in Fig. 5. 

Figure 6 shows the location of all variables with known colors on the cluster 
color-magnitude diagram (CMD). 
For the RR~Lyr stars marked positions correspond to the intensity-averaged
magnitudes. For the remaining variables we marked positions corresponding 
to the magnitude at maximum light. 
All but one RR~Lyr stars are grouped around $V\approx 19.5$
indicating that they belong to the SMC. RR~Lyr variable OGLEGC223 is
a background object in the galactic halo. 

There are 12 certain eclipsing binaries in our sample of variables. 
This group of stars is dominated by contact binaries with EW-type light
curves and periods shorter than 0.4 day. 
The only 3 stars whose light curves indicate a detached or semi-detached 
configuration are OGLEGC228, OGLEGC240 and OGLEGC253.  
OGLEGC240 is a detached binary with an EA-type light curve. 
The light curve of this variable is 
relatively noisy due to the faintness of the object. None the less
examination of the individual frames leaves no doubts about the reality of the
observed changes. The blue color and apparent magnitude of OGLEGC240 
indicates that it is an A spectral type  binary in the SMC. \\
OGLEGC228 shows a light curve typical of semi-detached binaries. 
This star is located among candidate blue-stragglers on the 
cluster CMD. OGLEGC253 is also a potential blue straggler.
Its light curve shows two minima of very different depth but
we cannot exclude possibility that the components of this binary are 
in geometrical contact.  Several systems with light curves similar to 
the light curve
of OGLEGC253 were analyzed during last decade (eg. Hilditch, King \&
McFarlane 1989).
Although most of detected binaries
are candidate blue stragglers, there are four contact systems
located slightly to the red of the cluster main sequence. 
These four binaries are potential main sequence systems belonging to 
47~Tuc. We shall return below to the question about membership of
identified contact binaries. 

Variables which could not be classified as either RR~Lyr stars 
or eclipsing binaries are generally red stars with periods ranging from 
2 days to several weeks. Six red variables which are located on or near 
the subgiant branch of 47~Tuc can be considered candidates for cluster members.
Recently Edmonds \& Gilliland (1996) reported discovery of low amplitude
variability among a large fraction of K giants in 47~Tuc.
Using the data collected with the HST they estimated that most of
variable giants have periods between 2 and 4 days and $V$ amplitudes 
in the range 5--25 mmag. Edmonds \& Gilliliand (1996) argue
that the observed variability of K giants from 47~Tuc is caused by
low-overtone pulsations. The variable K giants from our sample
have periods ranging from 2 to 36 days and show full amplitudes in the
$V$ band ranging from 0.08 to 0.18 mag. Based on the quality of our data we
estimate conservatively that we should be able to detect any periodic
variables among cluster giants with periods up to 2 weeks and full 
amplitudes exceeding 0.05 mag. We note that six candidates for 
variable K giants identified by us can easily be studied
spectroscopically.  Such observations would answer the question about 
the mechanism of observed photometric variability. Since observed light
variations are sufficiently large to imply detectable changes
of $V_{rad}$ if the variability is indeed due to pulsations. 

Variables with $V-I>1.1$ and $V>15.5$ are likely to be evolved stars on the
AGB in the SMC. We note that SMC stars can be easily distinguished from
47~Tuc members based on their radial velocities (heliocentric 
radial velocities of SMC and 47~Tuc 
are $+175$  km/s and $-18.7$ km/s, respectively).

We consider some of our period determinations as preliminary. 
Particularly, for OGLEGC229 we adopted $P=8.38$~$d$ because the 
light curve seems to show two distinct minima. However, we cannot
exclude the possibility that the correct period is in fact half this value. 
Also the period  of OGLEGC240 can be half the
adopted value of $P=4.32 d$. For $P=2.16$~$d$ our light curve of
OGLEGC240 would show just one detectable eclipse. 
\subsection{Cluster membership of the contact binaries}
The 47~Tuc cluster is located at a hight galactic
latitude of $b=-45$~deg. However, we cannot assume that 
all eclipsing binaries listed in Table 6 are cluster members. 
In particular, faint contact binaries with $V>16$ are  
known to occur at high galactic latitudes (eg. Saha 1984).
We have applied the absolute brightness calibration established by 
Rucinski (1995) to calculate $M_{\rm V}$ for the newly discovered 
contact binaries. Rucinski's calibration gives $M_{\rm V}$ as a function 
of period, unreddened color $(V-I)_{0}$ and metallicity:
\begin{eqnarray}
M_{\rm V}^{cal}=-4.43log(P)+3.63(V-I)_{0}
-0.31-0.12\times [{\rm Fe/H}].
\end{eqnarray}
We adopted for all systems $[{\rm Fe/H}]=-0.76$ and $E(V-I)=0.05$ 
(Harris 1996). 
Figure 7 shows the period versus an apparent distance modulus
diagram for contact binaries identified  in fields 104A-E.  
An apparent distance modulus was calculated for each 
system as a difference between its $V_{max}$ magnitude and $M_{\rm V}^{cal}$. 
An apparent distance modulus for 47~Tuc is estimated at     
$(m-M)_{\rm V}=13.21$ (Harris 1996). The only system with significantly 
deviating value of $(m-M)_{\rm V}$ is OGLEGC245.
This binary is most  probably a foreground variable. 
The remaining 8 systems plotted in Fig. 7 are likely members of the cluster.
\subsection{Completeness of the survey for contact binaries}
Our survey resulted in the identification of 8 contact binaries which are 
likely members of the cluster and 2 detached/semidetached
binaries which are possible blue stragglers belonging to the cluster.
Only 4 contact systems were identified below the cluster turnoff.
These numbers are surprisingly small considering that we
analyzed the light curves of 76119 stars with average magnitudes
$V<19.5$, mostly  main
sequence stars belonging to the cluster. For the clusters members the 
limiting magnitude $V=19.5$ corresponds to $M_{\rm V}=6.1$. 
We adopted here $(m-M)_{\rm V}=13.4$ for the apparent distance
modulus of 47~Tuc (Hesser et al. 1987).
The quality and quantity of photometry was sufficient to allow the detection
of potential eclipsing binaries with periods shorter than 1 day
and exhibiting eclipses deeper than about 0.3 mag (see Tables 1 \& 3).

A hint that our survey is quite complete with respect to
faint short period variables comes from the fact that we detected 12
RR~Lyr stars from the SMC. Graham (1975) searched for variables 
a field covering an area $1\deg \times 1.3 \deg$. His field was centered
north of 47~Tuc and included  a small part of the cluster. 
Graham identified 76  RR~Lyr stars, with  surface
density of 0.016 variables per arcmin$^{2}$. The effective 
area covered by our survey was 935 arcmin$^{2}$ yelding surface
density of RR~Lyr stars of about 0.013 variables per arcmin$^2$.
Apparently we did not miss in our survey too many RR~Lyr stars from the SMC.

The relative frequency of occurrence of $detectable$ contact binaries in 
our sample is $f_{c}=8/76119\approx 1.0E-4$. This frequency is more
than an order of magnitude lower than the binary frequency observed for fields 
containing galactic open clusters (Kaluzny \& Rucinski 1993; 
Mazur, Krzeminski \&
Kaluzny 1995) and for fields
located near the galactic center which were monitored by OGLE (Rucinski
1997). Recent surveys of globular clusters M71 (Yan \& Mateo 1994) 
and M5 (Yan \& Reed 1996) gave $f_{c}=4/5300\approx 7.5E-4$ and  
$f_{c}=5/3600\approx 1.4E-3$, respectively.  

To get a quantitative estimate of the completeness of
our sample we performed  tests with 
artificial variables for fields 104B and 104E.
Results of test for field 104B should apply
also to the fields 104A and 104C because all three fields contain similar 
numbers of measurable stars and were observed with comparable 
frequency. Similarly, results for field 104E should apply to field 104D.
For both fields we selected 5 samples
of objects from sets of stars whose light curves were examined for
variability. The brightest sample included stars with $16.0<V<17.0$ and
the faintest sample included  stars with $19.0<V<19.5$. a total of 100 stars
were selected at random from each sample. 
The observed light curves of these stars were then interlaced with
the synthetic light curves of model contact binaries.
The synthetic light curves were generated using
a simple prescription given by Rucinski (1993).
Two separate cases were considered. Case~I -- a  contact binary with
the inclination $i=60\deg$ and the mass ratio $q=0.10$.
Case~II -- a  contact binary with
the inclination $i=70\deg$ and the mass ratio $q=0.30$.
In both cases the so called "fill-out-parameter" was set to $f=0.5$.
The light curves corresponding to Case-I and Case-II show depths of
primary
eclipses  equal to 0.15 and 0.32 mag, respectively.
For  each of the artificially generated light curves a period was drawn 
in a random way from the range 0.2-0.45~d. Also the phase for the first
point of the given light curve was randomly selected.
The simulated light curves were then analysed in the manner as the
observed light curves.   
Specifically, we applied a procedure based on the  $\chi^{2}$ test.
The number of artificial variables which were "recovered"
for Cases I-II and 5 magnitude ranges is given in Table 7.
It may be concluded that the completeness of our sample of
contact binaries is  better than 88\% for systems with $V<19.5$ and 
depth of eclipses higher than 0.32 mag. For systems with full amplitudes
as small as 0.15 mag the completeness is higher than 73\% for $V<19.0$.
 
It has been noted by Kaluzny et al. (1997c) that the frequency of
occurrence of contact binaries in 47~Tuc is very low in comparison with 
open clusters and with several globular clusters which have been recently 
surveyed for eclipsing binaries by various groups. 
However, results presented here are based on a larger sample of
stars than the sample analyzed by Kaluzny et al. (1997c). 
A more extended discussion of this topic is given in Kaluzny et al.
(1997c). 
It is appropriate to note at this point that the low frequency of occurrence 
of contact binaries among 47~Tuc stars was first suggested
by Shara et al. (1988).
\section{The color-magnitude diagrams}
As a by product of our survey we obtained $V$ vs. $V-I$ CMD's
for all 5 monitored fields.
In Fig. 8 we show the CMD's for fields 104A and 104E. 
For each field the final photometry was obtained by merging measurements 
extracted from "long" and "short" exposures. Photometry 
obtained for fields 104A-B extends to brighter magnitudes than
photometry obtained for fields 104C-E. 
The frames used for construction of CMD's of monitored fields
are listed in Table 8.  Any detailed analysis of these data
is beyond the scope of this paper. 
We note only that our data can be used to select candidates for
cluster blue stragglers.

%
All photometry presented in this section was submitted in tabular form 
to the editors of A\&A and is available in electronic form to all
interested readers (see Appendix A). The potential users of this photometry
should be aware about possibility of some systematic errors of 
the photometry. These errors are most likely to be significant for 
relatively faint stars. The CCD chip used for observations by the OGLE 
suffers from some nonlinearity. 
More details on this subject can be found in Paper I. 
\section{Summary}
The main result of our survey is the identification of 8 contact binaries
which are likely members of 47~Tuc and 2 detached/semidetached
binaries which are possible blue stragglers. 
Particularly interesting is the bright binary OGLEGC228. By combining
radial velocity curves with photometry one would be able
to determine an accurate distance to this system. That would in turn
give distance to the cluster if the binary is indeed member of 47~Tuc. 
We failed to identify any detached eclipsing systems among cluster 
turnoff stars. Three such systems with periods ranging from 
1.5 to 4.6 day were identified in our survey of $\omega$~Cen (Papers I\&II).
 
We identified 6 variables which are likely to be red giants 
belonging to the cluster. These stars exhibit modulation of
luminosity with periods ranging from 2 to 36 days and full amplitudes
in the $V$ band ranging from 0.08 to 0.18 mag. They may represent
high-amplitude counterparts of low-amplitude variable K giants 
identified in the central region of 47~Tuc by Edmonds \& Gilliliand
(1996).

\begin{acknowledgements}
This project was supported by NSF grants AST-9530478 and AST-9528096
to Bohdan Paczynski. 
JK was supported also by the Polish KBN grant 2P03D-011-12.
We are indebted to Kyle Cudworth for sending us the astrometric
data on 47~Tuc. We thank Ian Thompson for his detailed remarks on the 
draft version of this paper. 
\end{acknowledgements}
\section{Appendix A}
Tables containing light curves of all variables discussed in this
paper as well as tables with $VI$ photometry for the surveyed fields
are published
by A\&A at the centre de Donn\'{e}es de Strasbourg, where they are available
in electronic form: See the Editorial in A\&A 1993, Vol. 280, page E1.
%

%
\clearpage
\setcounter{table}{0}
\begin{table}
\caption[]{Summary of observations collected for fields 104A-E.
$N_{\rm V}$ is the number of $V$-band images included in the data bases.}
\begin{flushleft}
\begin{tabular}{lll}
\hline\noalign{\smallskip}
Field & $N_{\rm V}$ & Dates of\\
      &     & observations  \\
\hline\noalign{\smallskip}
104A & 286 & Jun 17 - Sep 7, 1993 \\
104B & 270 & Jun 17 - Sep 7, 1993 \\
104C & 288 & Jun 16 - Sep 15, 1994  \\
104D & 125 & Jun 8 - Aug 22, 1995 \\
104E & 120 & Jun 8 - Aug 22, 1995 \\
\hline\noalign{\smallskip}
\hline\noalign{\smallskip}
\end{tabular}
\end{flushleft}
\end{table}
\setcounter{table}{1}
\begin{table}
\caption[]{Equatorial coordinates for the centers of fields 104A-E.}
\begin{flushleft}
\begin{tabular}{lll}
\hline\noalign{\smallskip}
Field & RA(1950) & DEC(1950) \\
      & h:m:s & deg:$\arcmin$ : $\arcsec$ \\
\hline\noalign{\smallskip}
104A & 0:19:52.7 & -72:22:45 \\
104B & 0:23:53.1 & -72:21:01 \\
104C & 0:21:47.9 & -72:10:35 \\
104D & 0:20:14.7 & -72:31:14 \\
104E & 0:23:10.4 & -72:31:22 \\
\hline\noalign{\smallskip}
\hline\noalign{\smallskip}
\end{tabular}
\end{flushleft}
\end{table}
\clearpage
\setcounter{table}{2}
\begin{table}
\caption[]{
Basic statistical data for stars in fields 104A-E examined 
for variability. The data are given in bins 0.5 mag wide.
Columns 2, 4, 6, 8 and 10 give {\it median} value of {\it rms} for
a given bin. Columns 3, 5, 7, 9 and 11 give the numbers of stars examined 
for variability.
}
\begin{flushleft}
\begin{tabular}{ccrcrcrcrcr}
\hline\noalign{\smallskip}
  &104A  &   & 104B  &    & 104C & & 104D & & 104E &    \\
V     &$<rms>$&  N  &$<rms>$& N  &$<rms>$& N  &$<rms>$& N &$<rms>$&N \\
\hline\noalign{\smallskip}
14.25 & 0.014&  214 & 0.014& 196 & 0.021& 100 &0.016&  112&0.012& 86 \\
14.75 & 0.017&  118 & 0.015& 112 & 0.014& 65  &0.012&   54&0.011& 69\\
15.25 & 0.016&  109 & 0.015& 115 & 0.014& 62  &0.012&   63&0.013& 57\\
15.75 & 0.015&  153 & 0.016& 165 & 0.012& 117 &0.017&   93&0.012& 67\\ 
16.25 & 0.018&  240 & 0.019& 244 & 0.016& 152 &0.018&  122&0.017& 111\\   
16.75 & 0.026&  434 & 0.023& 499 & 0.018& 261 &0.019&  222&0.020& 213\\  
17.25 & 0.027& 2057 & 0.027& 2181& 0.022& 1273&0.023& 1064&0.022& 1017\\   
17.75 & 0.032& 3149 & 0.032& 3248& 0.025& 1922&0.026& 1734&0.026& 1584\\   
18.25 & 0.041& 3973 & 0.039& 4089& 0.031& 2486&0.034& 2292&0.033& 2179\\   
18.75 & 0.052& 4574 & 0.054& 4884& 0.042& 3193&0.044& 2697&0.044& 2622\\   
19.25 & 0.072& 4515 & 0.071& 5000& 0.055& 3746&0.058& 3029&0.061& 2984\\  
19.75 & 0.102& 3803 & 0.100& 4248& 0.076& 3727&0.081& 2537&0.086& 2930\\   
20.25 & 0.148& 2986 & 0.144& 3029& 0.114& 3456&0.119& 2118&0.127& 2579\\  
\hline\noalign{\smallskip}
\hline\noalign{\smallskip}
\end{tabular}
\end{flushleft}
\end{table}
\clearpage
\setcounter{table}{3}
\begin{table}
\scriptsize
\caption[]{Rectangular and equatorial coordinates for variables
identified in the field of 47~Tuc. The X and Y coordinates give
positions of the variables on the template images (see text for details). 
 }
\begin{flushleft}
\begin{tabular}{lrrrrrr}
\hline\noalign{\smallskip}
Name & X & Y & RA(1950) & Dec(1950) & Field \\
     &   &   &  h:m:s & deg:$\arcmin$ : $\arcsec$ &       \\
\hline\noalign{\smallskip}
OGLEGC212 &  220.9 & 254.7 & 0:18:41.83 &-72:28:44.7& A\\
OGLEGC213 &  226.8 &1488.2 & 0:18:33.61 &-72:19:49.3& A\\
OGLEGC214 &  993.9 &1581.3 & 0:19:45.95 &-72:18:42.8& A\\
OGLEGC215 & 1338.7 & 541.9 & 0:20:26.80 &-72:26:01.4& A\\
OGLEGC216 & 1364.1 &1905.1 & 0:20:18.63 &-72:16:09.3& A\\
OGLEGC218 & 1706.3 &1622.4 & 0:20:53.31 &-72:17:59.4& A\\
OGLEGC219 &  344.5 & 244.3 & 0:22:54.32 &-72:27:00.8& B\\
OGLEGC220 &  206.5 &1776.8 & 0:22:30.21 &-72:16:00.3& B\\
OGLEGC221 &  390.8 &1673.4 & 0:22:48.44 &-72:16:39.2& B\\
OGLEGC222 &  643.2 & 280.3 & 0:23:22.66 &-72:26:35.0& B\\
OGLEGC223 &  768.9 & 816.9 & 0:23:30.69 &-72:22:37.9& B\\
OGLEGC225 &  709.6 &1398.1 & 0:23:20.72 &-72:18:27.9& B\\
OGLEGC226 &  931.0 &1853.5 & 0:23:38.38 &-72:15:02.5& B\\
OGLEGC227 & 1014.1 &  90.7 & 0:23:59.63 &-72:27:44.3& B\\
OGLEGC228 & 1041.4 & 657.1 & 0:23:57.90 &-72:23:37.6& B\\
OGLEGC229 & 1023.4 &1208.5 & 0:23:52.00 &-72:19:39.2& B\\
OGLEGC230 & 1131.5 &1315.2 & 0:24:01.46 &-72:18:49.1& B\\
OGLEGC231 & 1070.9 &1853.6 & 0:23:51.64 &-72:14:57.5& B\\
OGLEGC232 & 1512.2 & 176.1 & 0:24:46.64 &-72:26:49.3& B\\
OGLEGC233 & 1781.4 & 591.1 & 0:25:09.02 &-72:23:39.6& B\\
OGLEGC234 &  163.2 & 618.6 & 0:20:29.83 &-72:13:58.7& C\\
OGLEGC235 &  223.1 &1394.6 & 0:20:30.00 &-72:08:20.0& C\\
OGLEGC236 &  800.0 & 337.9 & 0:21:32.23 &-72:15:38.6& C\\
OGLEGC237 & 1403.4 &1232.5 & 0:22:22.42 &-72:08:48.9& C\\
OGLEGC238 & 1800.1 & 661.7 & 0:23:04.39 &-72:12:41.6& C\\
OGLEGC239 & 1359.7 & 528.5 & 0:22:23.80 &-72:13:55.8& C\\
OGLEGC240 & 1552.8 &1853.8 & 0:22:31.56 &-72:04:13.9& C\\
OGLEGC241 & 1649.5 & 992.6 & 0:22:47.51 &-72:10:23.8& C\\
OGLEGC242 &  130.6 & 969.0 & 0:18:49.43 &-72:32:03.6& D\\
OGLEGC243 &  328.9 & 931.3 & 0:19:08.78 &-72:32:14.1& D\\
OGLEGC244 & 1467.4 & 389.3 & 0:21:02.32 &-72:35:33.2& D\\
OGLEGC245 & 1227.3 &1183.7 & 0:20:33.52 &-72:29:56.4& D\\
OGLEGC246 & 1074.1 &1119.8 & 0:20:19.25 &-72:30:29.1& D\\
OGLEGC247 & 1559.2 & 285.6 & 0:21:11.95 &-72:36:15.1& D\\
OGLEGC248 & 1604.9 & 520.4 & 0:21:14.64 &-72:34:31.7& D\\
OGLEGC249 &  698.3 &  12.0 & 0:22:46.26 &-72:38:49.3& E\\
OGLEGC250 &  930.6 &1371.6 & 0:22:59.38 &-72:28:51.8& E\\
OGLEGC251 &  542.1 &1778.9 & 0:22:19.39 &-72:26:07.2& E\\
OGLEGC252 & 1171.0 & 909.6 & 0:23:25.73 &-72:32:04.6& E\\
OGLEGC253 & 1863.2 & 873.3 & 0:24:32.62 &-72:31:57.2& E\\
OGLEGC254 & 1629.1 &1072.4 & 0:24:08.63 &-72:30:38.8& E\\
OGLEGC255 & 1540.6 &1294.4 & 0:23:58.50 &-72:29:05.4& E\\
\hline\noalign{\smallskip}
\hline\noalign{\smallskip}
\end{tabular}
\end{flushleft}
\end{table}
\setcounter{table}{4}
\begin{table}
\caption[]{Light curve parameters for RR~Lyr stars
from the field of 47~Tuc. $A_{\rm V}$ is the full range of
variability.}
\begin{flushleft}
\begin{tabular}{cclll}
\hline\noalign{\smallskip}
Name    &     P  &$V-I$&  $V$ &$A_{\rm V}$ \\
OGLEGC&  day   &     & mean &    \\
\hline\noalign{\smallskip}
212   & 0.6946& 0.63& 19.5 & 0.8\\
213   & 0.6329& 0.56& 19.8 & 0.4\\
216   & 0.3617& 0.47& 19.9 & 0.4\\
223   & 0.2971& 0.33& 17.6 & 0.45\\
226   & 0.6474& ?   & 19.4 & 0.45\\
232   & 0.3635& 0.53& 19.5 & 0.5\\
234   & 0.6159& 0.79& 19.55& 0.6\\
235   & 0.5317& 0.42& 19.8 & 0.6\\
236   & 0.5083& 0.77& 19.8 & 0.3\\
243   & 0.6255& 0.56& 19.8 & 0.55\\
246   & 0.5719& 0.80& 19.65& 0.8\\
247   & 0.5115& 0.51& 19.9 & 0.65\\
255   & 0.5251& 0.50& 19.8 & 1.0\\
\hline\noalign{\smallskip}
\hline\noalign{\smallskip}
\end{tabular}
\end{flushleft}
\end{table}
%
\setcounter{table}{5}
\begin{table}
\caption[]{
Light-curve parameters for eclipsing binaries and 
red variables identified in the field of 47~Tuc.
Certain eclipsing systems and likely K giants belonging to the cluster
are marked in the second column.
}
\begin{flushleft}
\begin{tabular}{rllrrr}
\hline\noalign{\smallskip}
Name & Type  & Period& $V-I$ & $V{\rm max}$ & $V{\rm min}$ \\
OGLEGC&       & days  &              &               &        \\
\hline\noalign{\smallskip}
214&  Ecl& 0.2737& 0.82& 17.96 & 18.34\\
215&     & 8.666 & 1.14& 16.56 & 16.68\\
218&     & ?	  & 1.69& 15.80 & 16.17\\
219&  K  & 36.05  & 1.08& 15.28 & 15.46\\
220&  K  & 10.69 & 1.03& 16.265& 16.34\\
221&  Ecl& 0.3135& 0.79& 17.78 & 18.22\\
222&  K  & 18.93 & 0.95& 16.62 & 16.80\\
225&  Ecl& 0.2346& 1.04& 19.47 & 20.0\\
227&  Ecl& 0.3788& 0.52& 16.49 & 16.77\\
228&  Ecl& 1.1504& 0.34& 15.90 & 16.30\\
229&  K  & 8.378 & 1.06& 14.92 & 15.05\\
230&     & 4.814 & 1.23& 17.51 & 17.71\\
231&  K  & 6.498 & 0.93& 14.225& 14.325\\
233&     & 28.69  & 1.45& 16.55 & 16.72\\
237&  K  & 18.80  & 0.85& 16.87 & 16.95\\
238&  Ecl& 0.2506& 0.77& 18.46 & 18.80\\
239&     & ?	  & 1.53& 16.58 & 16.67\\
240&  Ecl& 4.3158 & 0.00& 19.93 & 20.65\\
241&     & ?	  & 1.67& 16.72 & 16.83\\
242&     & ?	  & 2.48& 16.55 & 17.42\\
244&  Ecl& 0.3837& 0.51& 16.16 & 16.38\\
245&  Ecl& 0.2789& 0.69& 15.49 & 15.87\\
248&     & 1.9967?& 1.26& 17.55:&  ?   \\
249&  Ecl& 0.3226& 0.64& 17.33 & 17.66\\
250&  Ecl& 0.3514& 0.43& 16.34 & 16.56\\
251&     & 3.4629 & 1.12& 16.56 & 16.87\\
252&     & ?	  & 2.90& 17.04 & 16.68\\
253&  Ecl& 0.4462& 0.57& 16.77 & 17.12\\
254&     & ?	  & 1.81& 16.47 & 16.62\\
\hline\noalign{\smallskip}
\hline\noalign{\smallskip}
\end{tabular}
\end{flushleft}
\end{table}
\clearpage
\setcounter{table}{6}
\begin{table}
\small
\caption[]{
Results of a test with artificial variables.
Columns 2-5 give numbers of recovered variables.
See text for details.}
\begin{flushleft}
\begin{tabular}{crrrr}
\hline\noalign{\smallskip}
Range    & Field 104B & Field 104B & Field 104E  & Field 104E \\
of $V$     & Case-I     & Case-II    & Case-I      & Case-II    \\
\hline\noalign{\smallskip}
16.0-17.0& 89   & 90 & 99 & 99 \\  
17.0-18.0& 88   & 94 & 90 & 94 \\
18.0-18.5& 81   & 96 & 89 & 94 \\
18.5-19.0& 74   & 89 & 73 & 92 \\
19.0-19.5& 52   & 88 & 35 & 88 \\
\hline\noalign{\smallskip}
\hline\noalign{\smallskip}
\end{tabular}
\end{flushleft}
\end{table}
\setcounter{table}{7}
\begin{table}
\caption[]{
List of frames used for construction of CMD's for
fields 104A-E.
}
\begin{flushleft}
\begin{tabular}{rrrcr}
\hline\noalign{\smallskip}
Frame & Field &$T_{\rm exp}$ & Filter & FWHM \\
      &       & sec          &        & arcsec \\
\hline\noalign{\smallskip}
mr5228 & 104A& 420  &V & 1.1 \\
mr5176 & 104A& 120  &V & 1.2 \\
mr8181 & 104A&	60  &V & 1.05  \\
mr5382 & 104A& 400  &I & 1.2 \\
mr5381 & 104A& 120  &I & 1.25 \\
mr8182 & 104A&	10  &I & 1.35 \\
mr5227 & 104B & 420 & V& 1.0\\
mr5177 & 104B & 120 & V& 1.4\\
mr8184 & 104B&	60  & V& 1.0\\
mr5385 & 104B&  400 & I& 1.3\\
mr5386 & 104B&  120 & I& 1.45\\
mr8183 & 104B&  10  & V& 1.0\\
mr7889 & 104C&  500 & V& 1.0\\
mr7902 & 104C&	50  & V& 1.0 \\
mr7903 & 104C & 500 & I& 1.0\\
mr7904 & 104C&	50  & I& 1.05\\
mr14597& 104D & 420 & V& 1.05\\
mr14589& 104D&	61  & V& 1.05 \\
mr14592& 104D&  420 & I& 1.20\\
mr14591& 104D&	60  & I& 1.15\\
mr14595& 104E&  420 & V& 1.1 \\
mr14596& 104E&	60  & V& 1.0\\
mr14593& 104E&  420 & I& 1.1\\
mr14594& 104E&   60 & I& 1.1\\
\hline\noalign{\smallskip}
\hline\noalign{\smallskip}
\end{tabular}
\end{flushleft}
\end{table}
\clearpage
Fig 1 -- A schematic chart showing location of fields 104A-E.
The cluster center is marked with a cross.
Each field covers $14.7\times 14.7$ armin$^{2}$.
North is up and east is to the left.
\vspace{0.4cm}
\newline
Fig. 2 -- Phased $V$ light curves for RR~Lyr stars from the SMC.
Inserted labels give the names of variables.
\vspace{0.4cm}
\newline
Fig. 3 -- Phased $V$ light curve for the halo RR~Lyr star OGLEGC223.
\vspace{0.4cm}
\newline
Fig. 4 -- Phased $V$ light curves for the variables
listed in Table 6. Inserted labels give the names of variables 
and their periods in days.
\vspace{0.4cm}
\newline
Fig. 4a -- Phased $V$ light curves for the variables
listed in Table 6. Inserted labels give the names of variables 
and their periods in days.
\vspace{0.4cm}
\newline
Fig. 5 -- Time domain light curves for variables with unknown
periods. Light curves for 1993 and 1994 seasons are shown for OGLEGC218.
\vspace{0.4cm}
\newline
Fig. 6 -- A schematic CMD for 47~Tuc with the positions
of the variables from fields A-E marked. The triangles represent certain
eclipsing binaries , the asterisks 
RR~Lyr stars and the open circles the remaining variables. 
Positions of stars from Table 6 are labeled. 
\vspace{0.4cm}
\newline
Fig. 7 -- Period vs. apparent distance modulus diagram for contact 
binaries from the field of 47~Tuc. A horizontal line at 
$(m-M)_{\rm V}=13.21$ corresponds to the distance modulus of the cluster.
Error bars correspond to the formal uncertainty in the  absolute magnitudes
derived using Rucinski's (1995) calibration.
\vspace{0.4cm}
\newline
Fig. 8 -- The CMDs for fields 104A (left) and 104E (right).
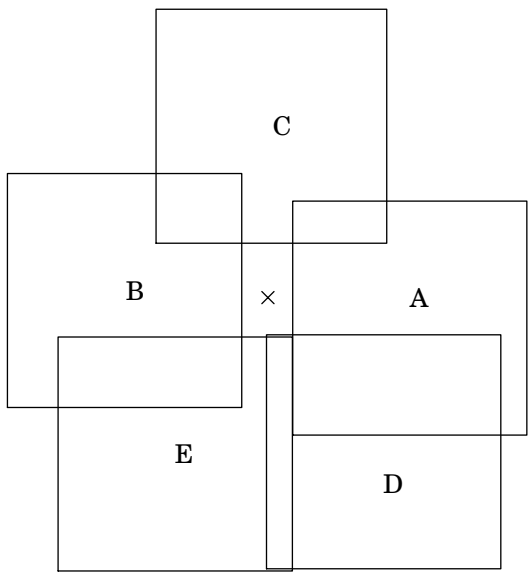
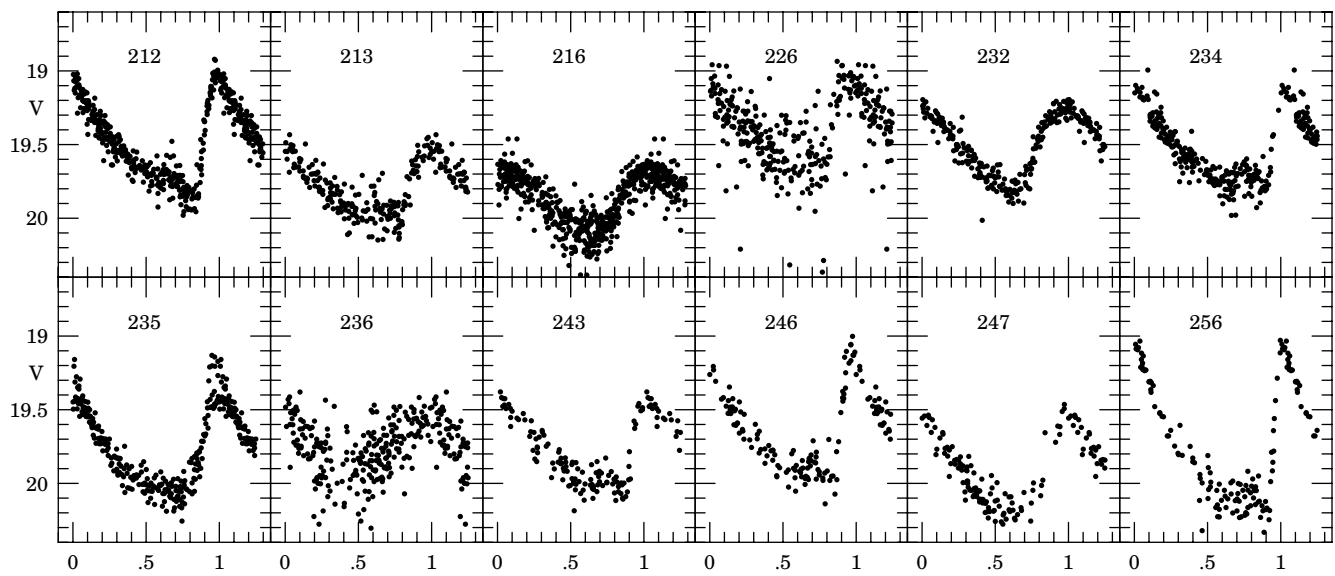
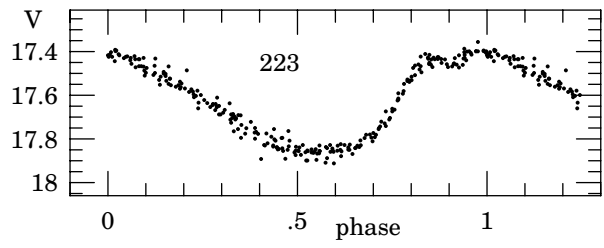
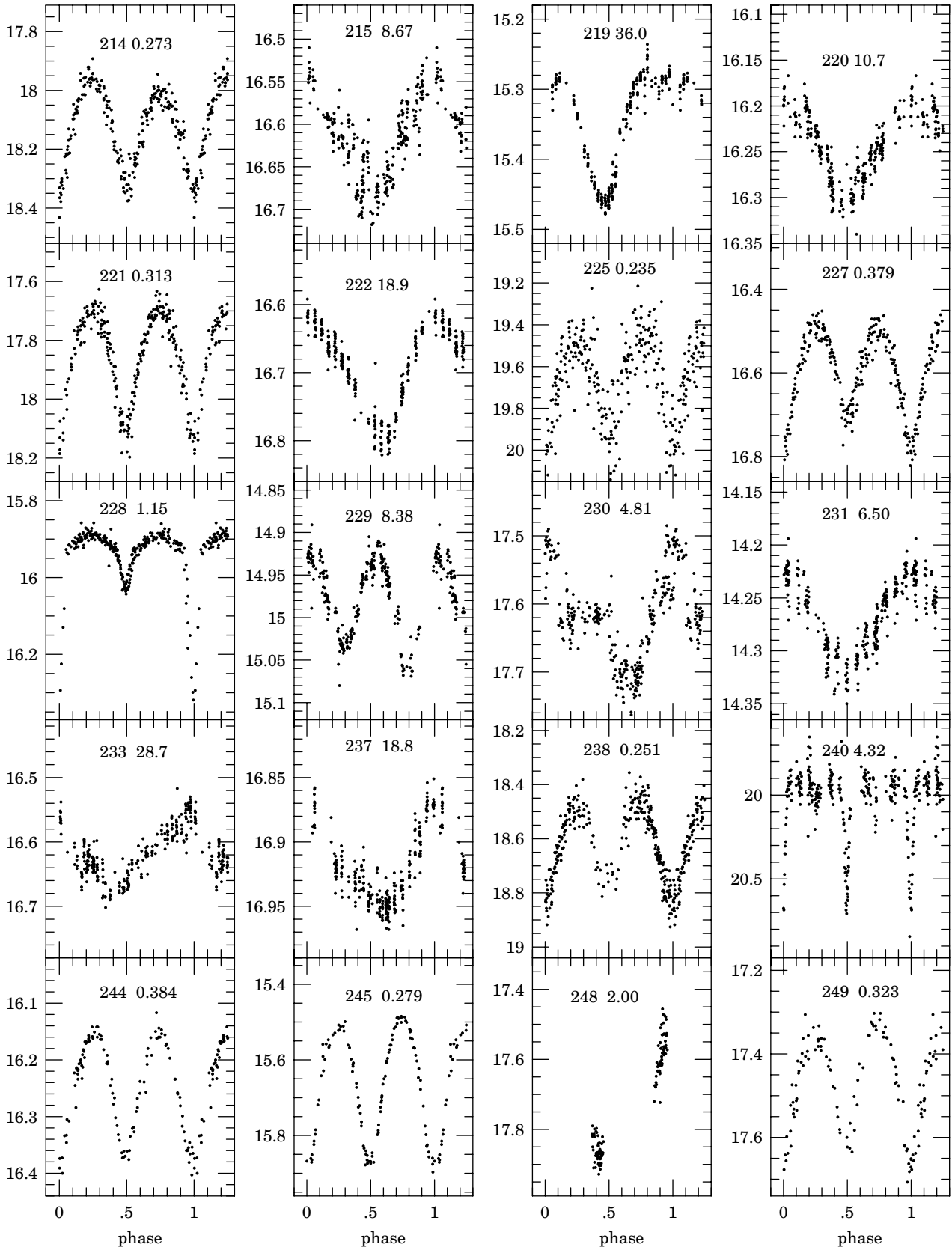
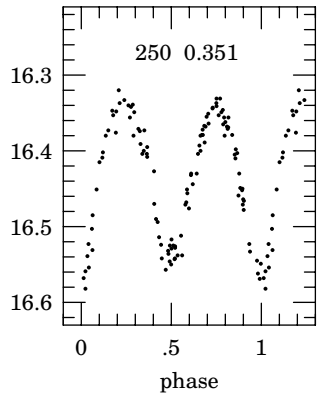
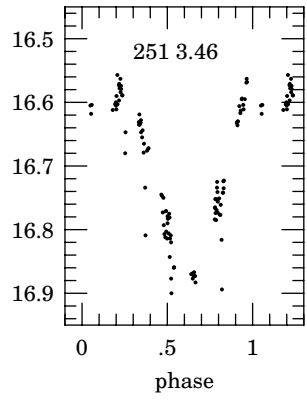
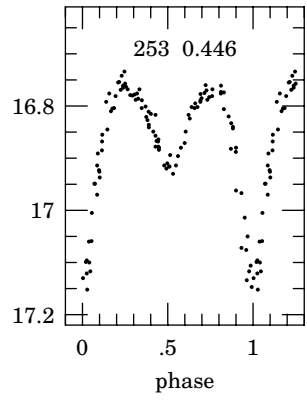
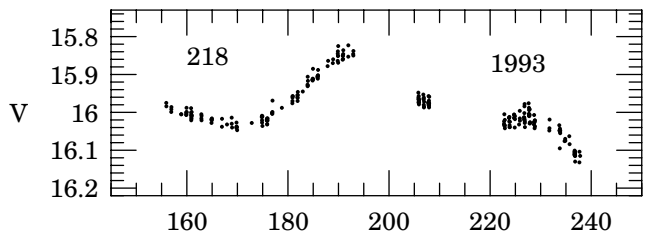
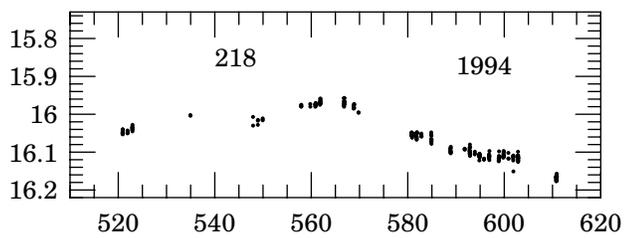
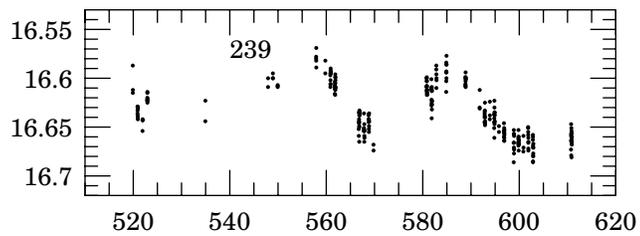
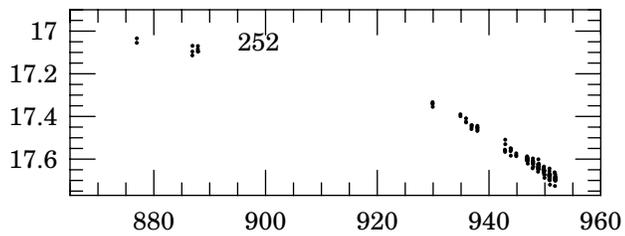
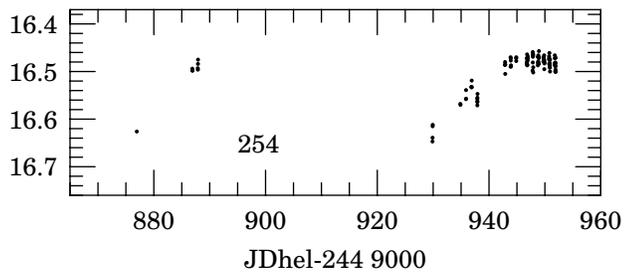
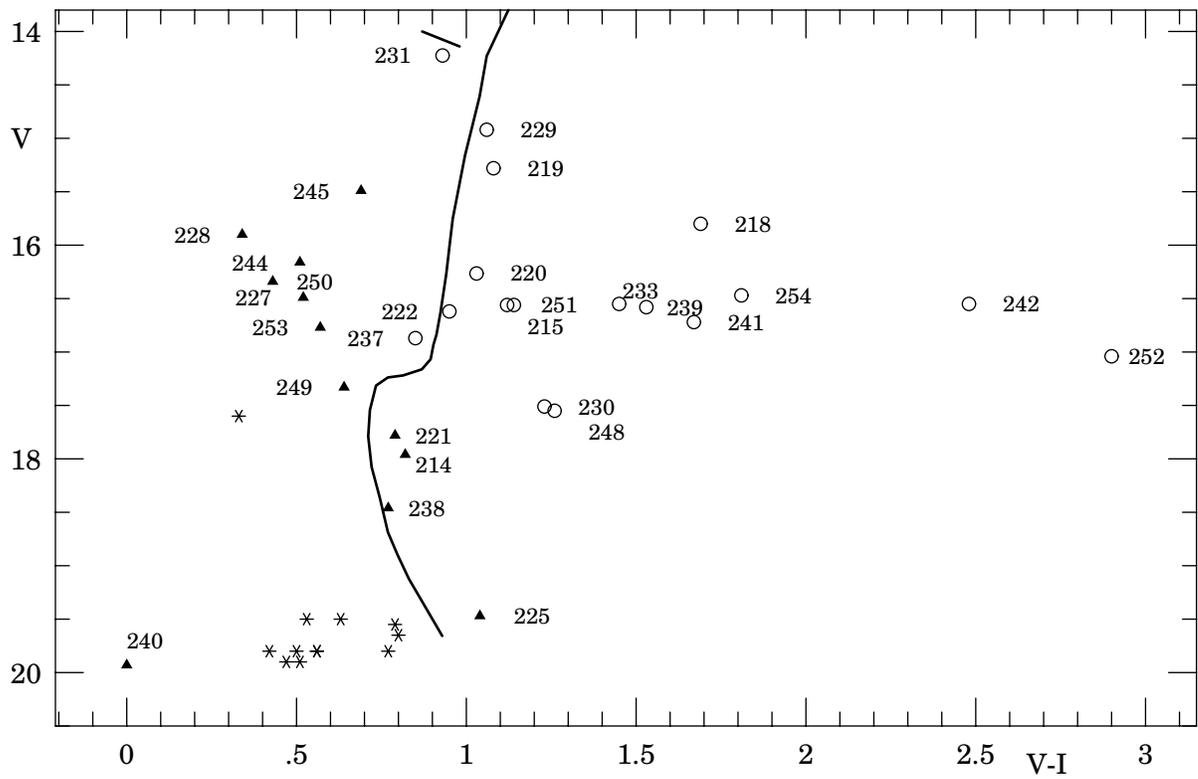
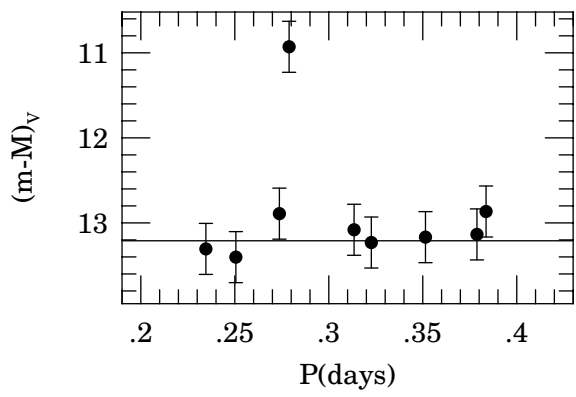
\end{document}